\documentclass[12pt]{iopart}
\usepackage{iopams}
\usepackage{graphicx}
\begin{document}

\title[High-field thermal transport properties of REBCO coated conductors]{High-field thermal transport properties of REBCO coated conductors}

\author{Marco Bonura and Carmine Senatore}

\address{Department of Applied Physics (GAP) and Department of Condensed Matter Physics (DPMC), University of
Geneva, quai Ernest Ansermet 24, CH-1211 Geneva, Switzerland}
\ead{marco.bonura@unige.ch}
\begin{abstract}

The use of REBCO coated conductors is envisaged for many applications, extending from power cables to high-field magnets. Whatever the case, thermal properties of REBCO tapes play a key role for the stability of superconducting devices. In this work, we present the first study on the longitudinal thermal conductivity ($\kappa$) of REBCO coated conductors in magnetic fields up to 19~T applied both parallelly and perpendicularly to the thermal-current direction.  Copper-stabilized tapes from six industrial manufacturers have been investigated. We show that zero-field $\kappa$ of coated conductors can be calculated with an accuracy of $\pm 15\%$ from the residual resistivity ratio of the stabilizer and the Cu/non-Cu ratio. Measurements performed at high fields have allowed us to evaluate the consistency of the procedures generally used for estimating in-field $\kappa$ in the framework of the Wiedemann-Franz law from an electrical characterization of the materials. In-field data are intended to provide primary ingredients for the thermal stability analysis of high-temperature superconductor-based magnets.

\end{abstract}

\maketitle

\section{Introduction}
High-temperature superconductors (HTS) have evolved considerably in the recent years and their use is presently envisaged in different branches of technology \cite{Senatore,Albiss,Jiang}. In particular, REBa$_2$Cu$_3$O$_{7-x}$ (REBCO, where RE = rare earth element) coated conductors (CCs) are considered of wide interest both for power applications and for magnet technology \cite{Albiss,Nguyen}. Many projects are being run for developing hybrid systems where HTS are used along with low-temperature superconductors (LTS) to generate magnetic fields well above $23.5$~T, which is the technological limit of Nb$_3$Sn-based magnets \cite{Markiewicz,Rey}. This would allow scientists to implement the present technology and to lay the foundations for next-generation HTS magnets.

The success in the development of HTS solenoids or multipoles relies on the control of electrical, mechanical and thermal properties of the superconductor \cite{Braccini,Turenne}. In spite of the large temperature margin, stability remains a key issue for HTS \cite{Iwasa}. In case of hybrid LTS-HTS magnets, irreversible damages of the HTS winding may occur in the case of a natural thermal runaway. In fact, when the LTS and HTS coils are inductively coupled, a quench of just the LTS part can provoke an abrupt rise of the current flowing in the HTS insert to values higher than the critical current with a consequent quench \cite{Maeda,Maeda2}. Quench detection in HTS windings is another major issue due to the very low values of the normal zone propagation velocity \cite{Lacroix,Lacroix2}.

Thermal conductivity ($\kappa$) of the CC is one of the most relevant parameters in the evaluation of the thermal stability of HTS-based devices. As far as we know, only few studies have been dedicated to thermal transport properties of CCs. Furthermore, all of them have been carried out in zero field \cite{Naito1,Naito2,Bagrets1,Bagrets2}. If a local heating of the tape occurs during operation, heat transfer through the conductor is the main channel for distributing heat and, consequently, preventing a quench. Longitudinal thermal conduction in technical superconductors (SCs) is ruled by the stabilizer, usually copper \cite{Wilson}. At cryogenic temperatures, $\kappa$ of Cu strongly depends on the intensity of the field as well as on the mutual orientation between magnetic field and thermal current \cite{Arenz,Simon}. Nowadays, the evaluation of the thermal stability of superconducting magnets is hindered by the lack of studies on the thermal properties of structural materials in high magnetic fields. Extensive investigations about magneto-thermal conductivity of Cu samples of different purity are limited to maximum fields of 8~T and have been carried out only with the field applied parallelly to the thermal current \cite{Simon}. However, electrical and thermal currents in a magnet are typically perpendicular to the direction of the magnetic field.

In practice, thermal conductivity of Cu even in high magnetic fields is deduced from electrical transport measurements in the frame of the Wiedemann-Franz law \cite{Manfreda}. Field-induced effects on the electrical resistivity of Cu have been reported in Ref.~\cite{Simon} for magnetic fields up to 10~T applied perpendicularly to the electric current. Relevant variations of the experimental results have been observed both for the magnetoresistance and for the Lorenz number as a function of sample properties (purity and geometry) and experimental conditions (magnetic field and temperature) \cite{Arenz,Simon,Abraham,White,Hust}.

In this work, we report on thermal conductivity of stabilized REBCO tapes at cryogenic temperatures in magnetic fields up to 19~T. This study provides the first direct comparison between actual in-field $\kappa$ values and estimated data as calculated from the the technical databases used for magnet design \cite{Manfreda}.

\section{Experimental Method}
Thermal transport properties have been investigated using an experimental setup specifically designed for high-field measurements in superconducting wires and tapes. The magnetic field has been applied in the plane of the tape both parallelly and perpendicularly to the thermal-current direction, the latter configuration being more relevant in the case of HTS solenoids. Heat power is supplied at one end of the sample and a gradient of temperatures is generated along the tape. A good thermal contact between sample and heater is ensured over the entire tape cross section. Thus, all the layers of the CC can contribute to the thermal transport. Once the steady-state heat flow is established, the temperature gradient along the sample, $\Delta T$, is measured by two Cernox bare chips directly glued on the sample surface using GE varnish. The two bare chips are calibrated during each measurement, allowing precise  temperature measurements. Heat power is set in order to have $\Delta T$ of the order of 0.1~K. The measurement uncertainty $\approx 5 \%$ is mostly due to inaccuracy in assessing the distance between the thermometers. Further details on the experimental setup are reported in Reference~\cite{Bonura1}.

Electrical resistivity measurements have been performed using a standard four-probe technique. The measurement uncertainty of $\approx 10 \%$ is mostly due to the determination of the sample dimensions and the distance between the voltage taps.

\section{Samples}\label{Samples}
We have investigated REBCO tapes produced by six manufacturers, namely American Superconductor (AMSC), Bruker HTS (BHTS), Fujikura, SuNAM, SuperOx and SuperPower. In the following, details on the tapes' layout are given only for layers potentially important for thermal conduction, either for the high $\kappa$ of the material or for the large cross-section fraction occupied. Layers' dimensions have been determined by optical microscopy.

Most of the manufacturers employ Hastelloy for the substrate, with the exception of AMSC and BHTS that use NiW and Stainless Steel, respectively \cite{Senatore}. The substrate thickness is $50~\mu$m for SuperPower, $60~\mu$m for SuNAM and SuperOx, $75~\mu$m for AMSC and Fujikura and $100~\mu$m for BHTS. For all tapes, a silver layer of about 1-2$~\mu$m is deposited over the REBCO layer. The tape from SuperPower includes a silver layer of about 1-2$~\mu$m also at the bottom of the substrate.

The examined CCs are thermally stabilized with Cu either by lamination or by electroplating. The first technique is adopted by AMSC and Fujikura. In particular, the CC from AMSC presents two Cu layers of about $50~ \mu$m, one on each tape face, whilst for the Fujikura tape only one $\approx 75~\mu$m-thick Cu strip is present atop the HTS side. For the other manufacturers, Cu is electroplated all around the tape, the total thickness of the stabilizer being $\approx30~\mu$m for BHTS, $\approx36$~$\mu$m for SuNAM, $\approx25$~$\mu$m for SuperOx and $\approx40$~$\mu$m for SuperPower. The thickness of the electroplated Cu layer is not the same on the two sides of the tape, the one next to the REBCO layer being thicker by 10-30\%, depending on the manufacturer. For the CC from BHTS the so-called ``dog bone'' shape has been observed, the tape thickness varying between $\approx 180$~$\mu$m at the edges and $\approx 135$~$\mu$m at the center. Table~\ref{TabSamples} summarizes some technical data of the CCs, namely the width ($w$) and the thickness ($t$) that define the tape cross-section dimensions, the material used for the substrate, the type of stabilization, and the cross-section fraction occupied by the copper ($s_{Cu}$).

In order to investigate the electrical resistivity of the stabilizer at low temperature, we have extracted Cu samples (length $\approx 4$~cm) from the CCs. In the case of laminated tapes, Cu strips have been unsoldered using a heating plate at 250$^{\circ}$C. The procedure lasts about 1~min, thus limiting possible annealing effects. Electroplated Cu has been isolated removing with sand paper the other deposited layers of the CC. Only the copper facing the substrate and the substrate itself are left with this procedure. This does not affect the measurement since the electrical resistivity of the materials used for the substrate is much larger than that of Cu \cite{Simon,Lu,Bauer}.

\begin{table}[t]
\caption{\label{TabSamples}Technical data of the investigated REBCO tapes (symbols defined in the text)}
\footnotesize\rm
\begin{tabular*}{\textwidth}{@{}l*{15}{@{\extracolsep{0pt plus12pt}}l}}
\br
Manufacturer& Dimensions ($w \times t$)&Substrate&Type of stabilization& $s_{Cu}$\\
\mr
AMSC&$4.8 \times 0.20$ mm$^2$&NiW&Cu laminated on 2 sides&0.51\\
BHTS&$4.1 \times 0.15$ mm$^2$&Stainless Steel&Cu electroplated&0.20\\
Fujikura&$5.1 \times 0.16$ mm$^2$&Hastelloy&Cu laminated on 1 side&0.44\\
SuNAM&$4.0 \times 0.11$ mm$^2$&Hastelloy&Cu electroplated&0.34\\
SuperOx&$4.0 \times 0.09$ mm$^2$&Hastelloy&Cu electroplated&0.27\\
SuperPower&$4.0 \times 0.10$ mm$^2$&Hastelloy&Cu electroplated&0.40\\
\br
\end{tabular*}
\end{table}
\section{Experimental Results}\label{Results}
Thermal conduction properties of REBCO CCs from different manufacturers have been investigated in the range of temperatures from $\approx 4$~K to $\approx 40$~K. Measurements have been performed at zero field and in magnetic fields up to 19~T with the field oriented both perpendicularly and parallelly to the thermal current. An extra data point has been measured at liquid-nitrogen temperature and zero field. In-field measurements have been performed at $B=1$~T, $B=7$~T and $B=19$~T in the case of thermal current perpendicular to the magnetic-field direction and at $B=19$~T with the thermal current parallel to the field. This has allowed us to evaluate the field-orientation effects on the thermal transport.

\begin{figure}[p]
\centering \includegraphics[width=16 cm]{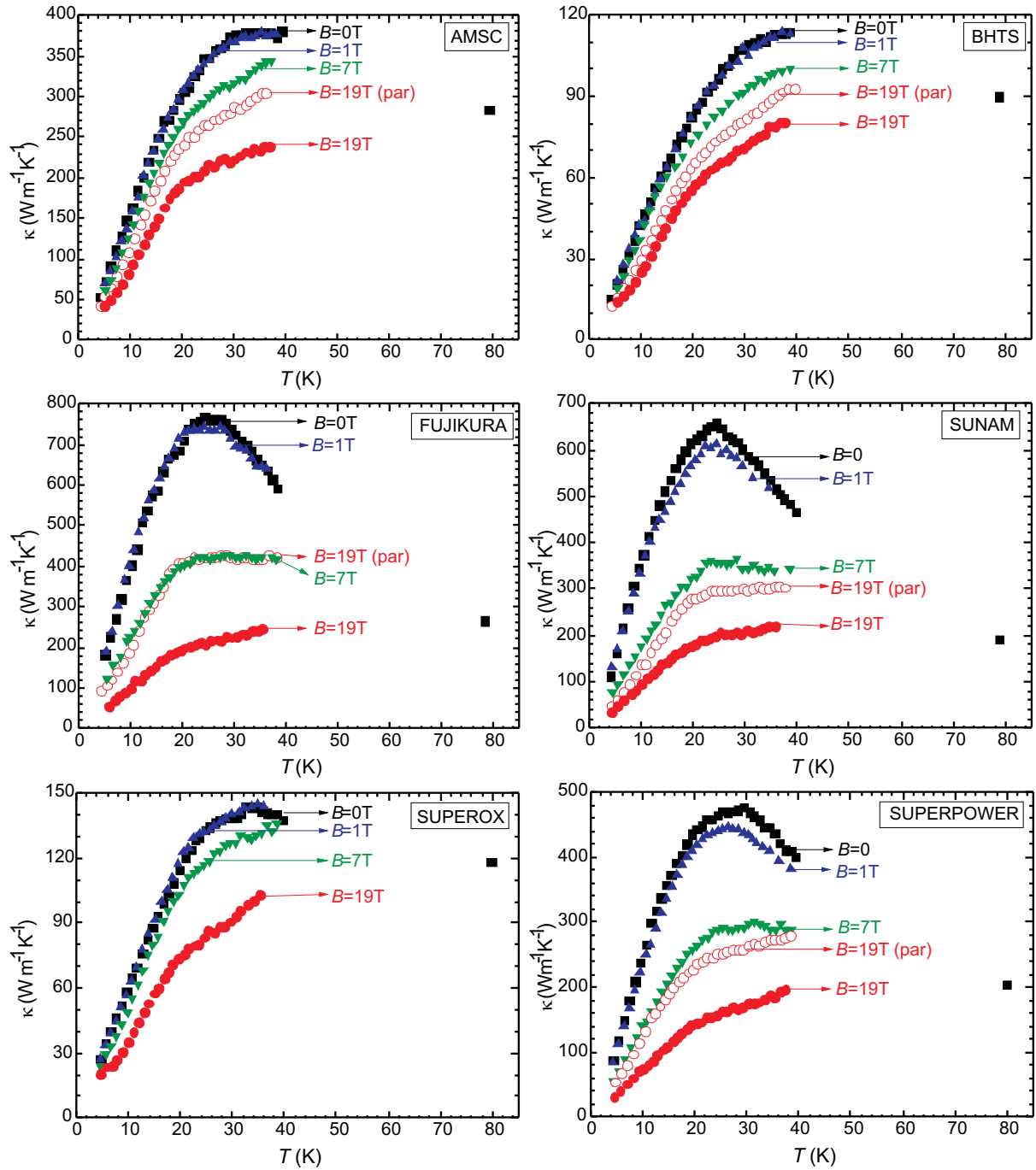} \caption{\label{All_k}
Temperature and magnetic-field dependence of the thermal conductivity in REBCO CCs from various manufacturers. Data points acquired with the magnetic field parallel to the thermal-current direction are shown as open symbols.}
\end{figure}

Figure~\ref{All_k} presents the experimental results for the examined CCs. The comparison shows that the $\kappa$($T$) curves at $B=0$ present a maximum whose position depends on the specific CC. The temperature value at which the maximum in the $\kappa(T,B=0)$ curve occurs, $T_{max}$, is reported in Table~\ref{TabSamples2} for each tape. The examined CCs exhibit very different zero-field $\kappa$ values at the maximum. $\kappa$ varies from $\approx 114$~Wm$^{-1}$K$^{-1}$ for Bruker HTS to $\approx 765$~Wm$^{-1}$K$^{-1}$ for Fujikura. $\kappa$ values at $T=T_{max}$ measured at $B=0$ and $B=19$~T in the transversal configuration are shown in Table~\ref{TabSamples2}.

For all the investigated samples, we have observed a reduction of $\kappa$ with increasing the field. Also the relative orientation between thermal-current and magnetic-field directions plays a role, the field-induced effects being more pronounced when the field is perpendicular to the heat flow. For the CC from Fujikura, $\kappa(T)$ values at $B=7$~T in the perpendicular configuration are very close to those measured at $B=19$~T with the field parallel to the thermal current. To perform a quantitative comparison of the field-induced effects on $\kappa$, we have calculated the relative reduction $\zeta(T,B)\equiv [\kappa(T,0$)-$\kappa(T,B$)]/$\kappa(T,0)$, with $B$ perpendicular to the heat-flow direction. The $\zeta(T_{max},B)$ values, expressed in percentage, are reported in Table~\ref{TabSamples2} for $B=1$~T , $B=7$~T and $B=19$~T.

\begin{figure}[p]
\centering \includegraphics[width=16 cm]{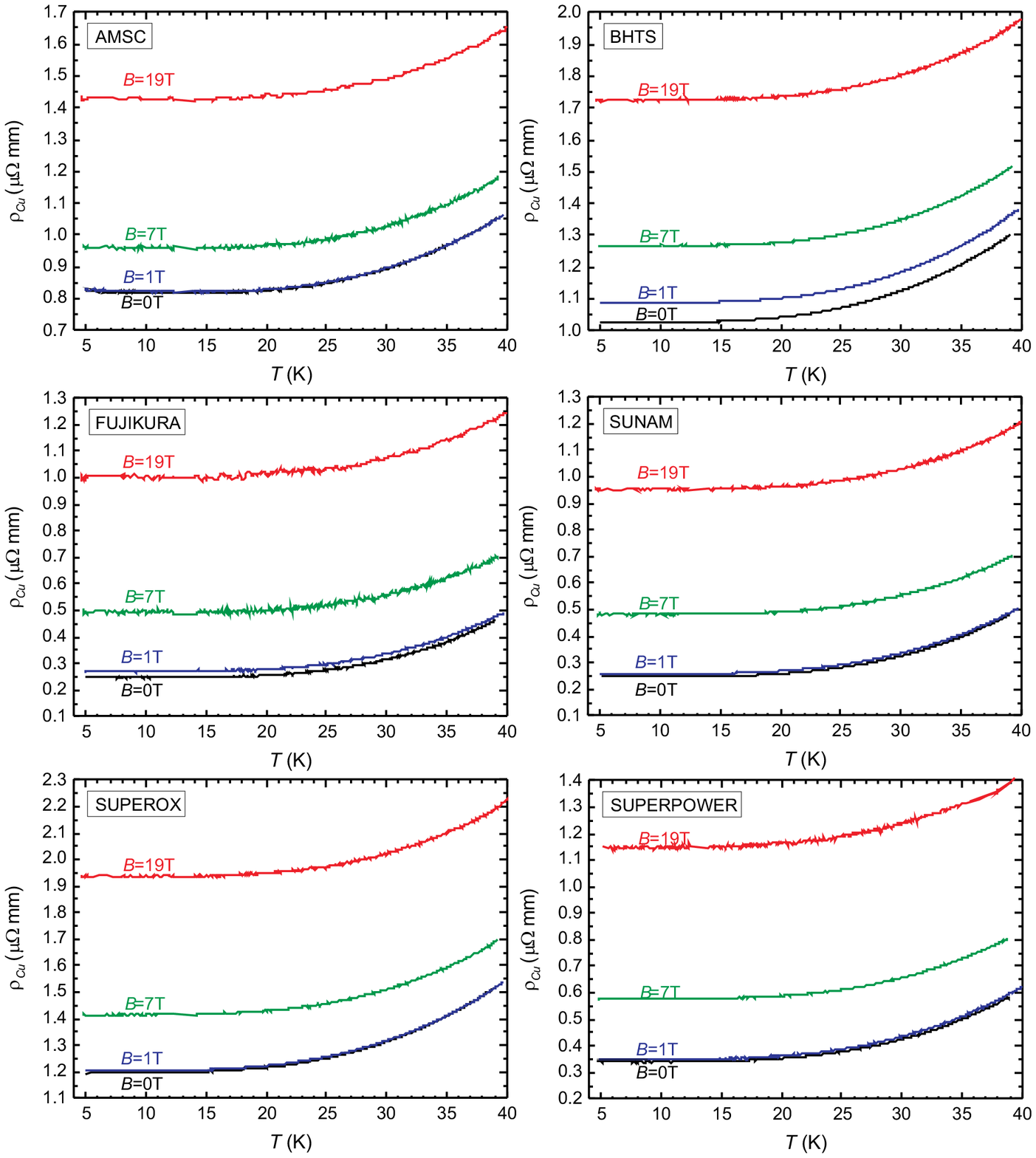} \caption{\label{All_R}
Temperature and magnetic-field dependence of the electrical resistivity as measured on Cu samples extracted from the CCs as detailed in the text. The magnetic field has been applied perpendicularly to the current direction.}
\end{figure}

In the case of stabilized CCs, it is expected that the longitudinal heat transport is ruled by the metal stabilizer \cite{Wilson}. In a metal, electron-defect scattering processes constitute the main channel for transmitting thermal energy at low temperature. As a consequence, an important parameter for modeling the thermal transport in metals is the residual resistivity ratio, $RRR\equiv \rho$(273~K)/$\rho_{res}$, where $\rho_{res}$ is the residual electrical resistivity measured at low temperature. In REBCO tapes, the $RRR$ of the Cu stabilizer cannot be determined just measuring the temperature dependence of the resistivity \cite{Bagrets1}. This is a consequence of the fact that for $T<T_C$ the electric current flows in the superconducting layer and this prevents the evaluation of $\rho_{res}$ for Cu. For circumventing this issue, we have measured the electrical resistivity on Cu samples extracted from the CCs as detailed in Section~\ref{Samples}. The results are shown in Fig.~\ref{All_R}. Measurements have been performed in the range of temperatures 4-40~K at zero field and at $B=1$~T, 7~T and 19~T, with the magnetic field perpendicular to the electrical current. Furthermore, the $\rho$ value at $B=0$ has also been measured at 273~K. The $RRR$ values as deduced from resistivity are reported in Table~\ref{TabSamples2}.

\begin{table}[t]
\caption{\label{TabSamples2}Thermal and electrical properties of the investigated coated conductors (symbols defined in the text)}
\footnotesize\rm
\begin{tabular*}{\textwidth}{@{}l*{15}{@{\extracolsep{0pt plus12pt}}l}}
\br
&AMSC&BHTS&FUJIKURA&SUNAM&SUPEROX&SUPERPOWER\\
\mr
$T_{max}$ [K]&34&38&25&24&34&28\\
$\kappa$($T_{max}$,0~T) [Wm$^{-1}$K$^{-1}$] & 380~ & 114 & 765 & 650 & 142 & 475\\
$\kappa$($T_{max}$,19~T) [Wm$^{-1}$K$^{-1}$] & 235 & 80 & 215 & 198 & 101 &169\\
$\zeta(T_{max}$,1~T) &  $<$5\% & $<$5\% & $<$5\% & 6\% & $<$5\% &6\%\\
$\zeta(T_{max}$,7~T) &  12\% & 12\% & 44\% & 44\% & 8\% &39\%\\
$\zeta(T_{max}$,19~T) &  38\% & 29\% & 71\% & 66\% & 29\%&64\%\\
$RRR_{Cu}$ from $\rho(T)$ & 19 & 17 & 59 & 61 & 14 &42\\
$RRR_{Cu}$ from $\kappa(T)$ & 20 & 14 & 62 & 69 & 13 &39 \\
$\Pi$(30K,19T) [$\mu$WK$^{-1}$m]&110&9&81&30&10&28\\
$J_{Eng}$($4.2$~K,19~T) [A/mm$^2$] & 75 & 290 & 305 & 220 & 445 & 345\\
\br
\end{tabular*}
\end{table}
%


\section{Discussion} \label{Discussion}
\subsection{Thermal Conductivity at B=0} \label{Discussion-k(0)}
Longitudinal thermal transport in composite conductors can be treated with a formalism analogous to the case of electrical resistances connected in parallel. The overall thermal conductivity ($\kappa_{tot}$) is the weighted sum of the thermal conductivity of each layer, the weight being the surface fraction $s_i \equiv S_i / S_{tot}$, where $S_i$ and $S_{tot}$ are the surface of the cross section occupied by the $i^{th}$ component and the total cross-section area of the conductor, respectively: $\kappa_{tot}=\sum \kappa_i s_i$.

\begin{figure}[t,b]
\centering \includegraphics[width=8 cm]{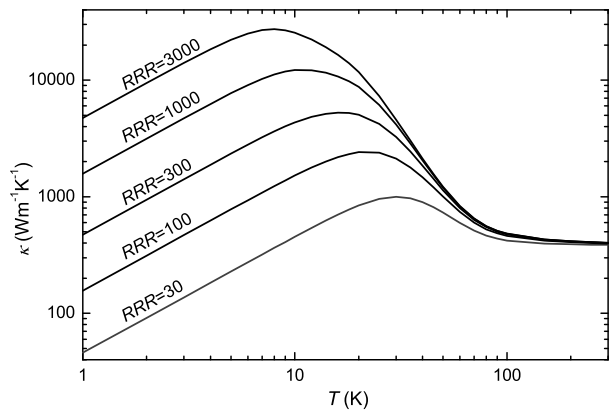} \caption{\label{k_Cu_Graph}
Temperature dependence of the thermal conductivity at zero magnetic field expected for copper with different $RRR$. Curves have been derived from Equation~~\ref{k_Cu}.}
\end{figure}

Thermal conductivity of the materials used for the substrate (NiW, Stainless Steel or Hastelloy), which occupies an important fraction of the tape section, is low when compared to that of copper (see Figure~\ref{k_Cu_Graph}), being $\kappa\lesssim10$~Wm$^{-1}$K$^{-1}$ at cryogenic temperatures \cite{Lu,Bauer,Bagrets3}. Also the contribution of the silver layer is negligible, due to its low cross-section area \cite{Smith}. Thus, the product $\kappa_{Cu} s_{Cu}$ at cryogenic temperatures is expected to be at least one order of magnitude higher than similar products for other constituents.

Thermal transport properties of Cu at zero field have been widely studied \cite{Simon,Hust}. It has been shown that $\kappa$($T$) values at $B=0$ can be estimated with an uncertainty of $\pm 15\%$ from the $RRR$ \cite{Hust,Rosemberg}. This is a consequence of the fact that for pure metals the electronic contribution to heat conduction is predominant with respect to the lattice one. In particular, the thermal conductivity of Cu can be expressed by:
\begin{eqnarray}\label{k_Cu}
    \kappa_{Cu}=(W_0+W_i+W_{i0})^{-1} ,\\
    W_0= \frac{\beta}{T} ,\nonumber \\
    W_i=\frac{P_1 T^{P_2}}{1+P_1 P_3 T^{(P_2+P_4)}e^{-(P_5/T)^{P_6}}} ,\nonumber \\
    W_{i0}=P_7\frac{W_i W_0}{W_i+W_0} , \nonumber
\end{eqnarray}
with $\beta\approx 0.634 / RRR$, $P_1 = 1.754 \times10^{-8}$, $P_2 = 2.763$,
$P_3 = 1102$, $P_4 = -0.165$, $P_5 = 70$, $P_6 = 1.756$, $P_7 =0.235 \cdot RRR ^{0.1661}$ in SI units \cite{Hust}. $W_0$ and $W_i$ represent the electron-defect and the electron-lattice scattering contributions, respectively. $W_{i0}$ is an interaction term between $W_0$ and $W_i$ \cite{Hust}.

Figure~\ref{k_Cu_Graph} shows the $\kappa$($T$) curves for Cu at various $RRR$ as calculated from Equation~\ref{k_Cu}. At low temperature, samples with higher purity have better thermal transport properties. This can be easily understood considering that thermal conductivity is proportional to the electron mean free path that in turn depends on the amount of defects present in the material \cite{Kittel}. The purity of the material becomes less important for $T\gtrsim 70$~K when the electron-phonon scattering dominates over the electron-defect one. This suggests that the quality of Cu used for stabilizing HTS is not essential in the case of use at liquid-nitrogen temperature, whilst it can be crucial for low-temperature applications, as in the case of high-field magnets.

The Wiedemann-Franz law quantitatively defines the correlation between the electrical and thermal conductivities in metals, being characterized by the Lorenz number, $L=\kappa\rho/ T$. At low temperatures, $L=L_0=\frac{\pi^2}{3} (\frac{k_B}{e})^2\approx2.44 \times 10^{-8}$~W$\Omega$K$^{-2}$, where $k_B$ is the Boltzmann constant and $e$ the electron charge \cite{Froehlich}. The derivation of $L_0$  does not require any specification about the shape of the Fermi surface, thus extending its validity to all metals. Predictions of the Wiedemann-Franz law are in good agreement with experiments at low temperatures and zero magnetic field, provided that the phonon contribution to thermal transport is negligible and that electron scattering processes are elastic \cite{Arenz,White}.

In Section~\ref{Results} we have pointed out that the position of the peak in the $\kappa(T)$ curve measured at $B=0$ depends on the sample properties. From Figure~\ref{k_Cu_Graph} one can observe that the position of the maximum is determined by the $RRR$. In particular, $T_{max}$ is lower when $RRR$ is higher. This can be understood in terms of the Wiedemann-Franz law. At very low temperatures, the electrical resistivity is constant being $\rho=\rho_{res}$. It follows that $\kappa$ is expected to increase linearly with $T$. On further increasing the temperature, $\rho$ starts increasing as $T^n$ with $n>1$ \cite{Borchi}, implying a consequent reduction of $\kappa$. As the temperature range where $\rho=\rho_{res}$ is wider for low $RRR$ values, it follows that dirtier samples have higher $T_{max}$.

\begin{figure}[t,b]
\centering \includegraphics[width=8 cm]{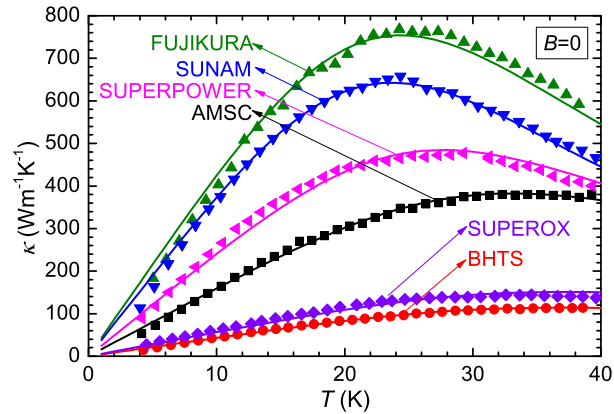} \caption{\label{Fig_k(T,0)}
Temperature dependence of the thermal conductivity of REBCO CCs from different manufacturers at $B=0$. The lines indicate the best-fit curves obtained as described in the text.}
\end{figure}

In Figure~\ref{Fig_k(T,0)}  we compare the $\kappa(T)$ curves measured at $B=0$ for all the examined CCs. Continuous lines are the best-fit curves obtained using Equation~\ref{k_Cu} with the $RRR$ of Cu as the only fitting parameter (assuming $\kappa_{tot}= \kappa_{Cu}s_{Cu}$). The $s_{Cu}$ values used for the fit have been determined from optical micrographs and have been reported in Table~\ref{TabSamples}. The agreement between experimental and best-fit curves is very good for all the investigated samples. The best-fit values for the $RRR$ of Cu are reported in Table~\ref{TabSamples2}. The discrepancy between $RRR$ values obtained by the best-fit procedure and those found from the electrical resistivity measurements is smaller than $15\%$ for most of the samples, thus within the accuracy of the interpolation formulas (Equation~\ref{k_Cu}) for Cu \cite{Hust}.

The consistency of the best-fit procedure indicates that $\kappa$ values at $B=0$ of REBCO CCs can be properly estimated from the $RRR$ of the copper stabilizer and the Cu/non-Cu ratio. Beyond the correspondence between electrical and thermal measurements, we have also evaluated the fluctuations of the $RRR$ in samples extracted from the same CC batch. We have found variations smaller than $10\%$ for all the manufacturers.

\subsection{In-field Thermal Conductivity} \label{Discussion-k(B)}
We are now interested to validate the procedures typically used to deduce in-field thermal properties from electrical measurements. Two comprehensive monographs from the US National Institute of Standards and Technology (NIST) report about in-field thermal resistance of Cu samples with different $RRR$ \cite{Simon,Hust}. Only measurements with fields up to 8~T applied parallelly to the thermal current are shown. A field-induced reduction of $\kappa$ is reported, the relative variation being higher for purer samples. This trend is confirmed by our measurements (see $\zeta$ values in Table~\ref{TabSamples2}).

Arenz \textit{et al.} have investigated thermal and electrical conductivities of a Cu wire with $RRR=108$ in magnetic fields up to 12.5~T, both in transverse and longitudinal orientations \cite{Arenz}. They have observed that $\kappa$ is more intensely affected by fields applied perpendicularly to the thermal current, in agreement with our findings. Furthermore, they have found that the Lorenz number is nearly independent of the magnetic field only in the parallel configuration, whilst it increases with the field in the perpendicular one.

\begin{figure}[t,b]
\centering \includegraphics[width=8 cm]{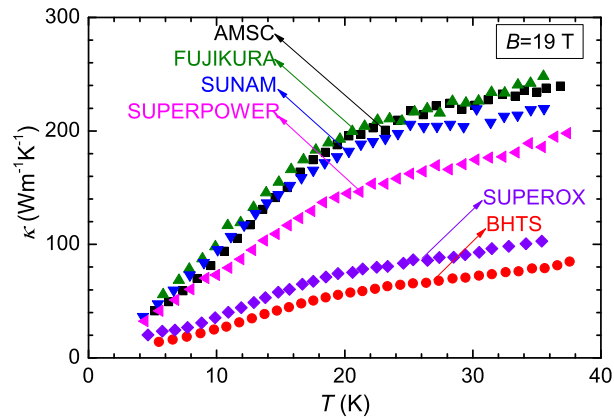} \caption{\label{Fig_k(T,19)}
Temperature dependence of the thermal conductivity of REBCO CCs from different manufacturers at $B=19$~T (field applied perpendicularly to the thermal-current direction).}
\end{figure}

In Figure~\ref{Fig_k(T,19)} we show the $\kappa(T)$ curves measured at $B=19$~T with the field perpendicular to the thermal current. Comparing the results shown in Figure~\ref{Fig_k(T,0)} and Figure~\ref{Fig_k(T,19)} we deduce that the differences in the thermal properties of the CCs observed at $B=0$ are strongly reduced after the application of an intense magnetic field. It is noteworthy that some CCs with very different $\kappa$ values at $B=0$ have comparable thermal conduction properties at $B=19$~T, as for the tapes from AMSC, Fujikura and SuNAM. This results from two causes: i) different relative reduction of $\kappa$ on applying the magnetic field; ii) different amount of Cu present in the tapes. In fact, the field-induced reduction of $\kappa$  in Cu depends on the $RRR$ and samples with higher purity show larger variations \cite{Simon}. Furthermore, since the Cu contribution to the longitudinal thermal conductivity of the CC is $\kappa_{Cu}s_{Cu}$, it follows that the measured $\kappa$ reduction depends also on the Cu/non-Cu ratio.

The field-induced variations of $\kappa$ can be qualitatively understood considering the reduction of the electron mean free path caused by the Lorenz-force action. If the field-induced scattering processes are elastic, both thermal and electrical conductivities are reduced by the same extent and consequently, the Lorenz number is not affected by the magnetic field. This is not the case if the scattering event implies a variation of the electron energy, since ``vertical movements'' on the Fermi surface have an effect only on the thermal conductivity \cite{White}. If the Lorenz number does not vary with the magnetic field, in-field $\kappa$ values can be inferred by measuring the magnetoresistance of the stabilization metal, since:
\begin{equation}\label{MR}
    \kappa(T,B)=\frac{L(T,B) T}{\rho(T,B)}\approx \frac{L(T,0) T}{\rho(T,B)} = \frac{\rho(T,0)}{\rho(T,B)} \kappa(T,0) \, .
\end{equation}

Transverse magnetoresistance of Cu has been widely investigated in the past. Simon \textit{et al.} \cite{Simon} have collected data in magnetic fields up to $B=10$~T from different references and have summarized them in a unique Kohler plot where the fractional increase in resistance $\Delta\rho$($T,B$)$\equiv[\rho(T,B)-\rho(T,0)]/\rho(T,0)$ is shown as a function of the product $B\times [\rho(273$~K,$B=0$)/$\rho(T$,$B=0$)]. The authors have derived an average Kohler curve that allows evaluating in-field electrical resistivity values starting from $\rho$ values measured at $B=0$. The uncertainty of the Kohler curve, estimated from the variance of the data, is also given \cite{Simon}.

Though the Kohler formula has been elaborated in a limited range of magnetic fields, the magnetoresistance values at $B>10$~T are commonly used in relevant software and databases of material properties \cite{Manfreda}. This allows calculating $\kappa$($T,B$) from Equations~\ref{k_Cu}-\ref{MR}, once measured the resistivity at $B=0$. Also the uncertainty of the Kohler plot is an important parameter for evaluating the stability margins of a SC-based device. In the case of REBCO CCs operating at 4.2~K, the temperature margin is typically $\approx25$~K, implying that the current sharing should occur at $T^*\approx 30$~K \cite{Iwasa}. For Cu with $RRR= 50$ one finds from the Kohler plot that $\Delta\rho$(30~K,19~T)$\approx 0.36 \pm 0.12$. This means that in-field $\kappa$ values can be estimated from Equation~\ref{MR} with an uncertainty of about $\pm 35\%$. If $\kappa(T,0)$ is calculated from the $RRR$ using Equations~\ref{k_Cu}, the overall uncertainty becomes $\pm 50\%$.
\begin{figure}[t,b]
\centering \includegraphics[width=8 cm]{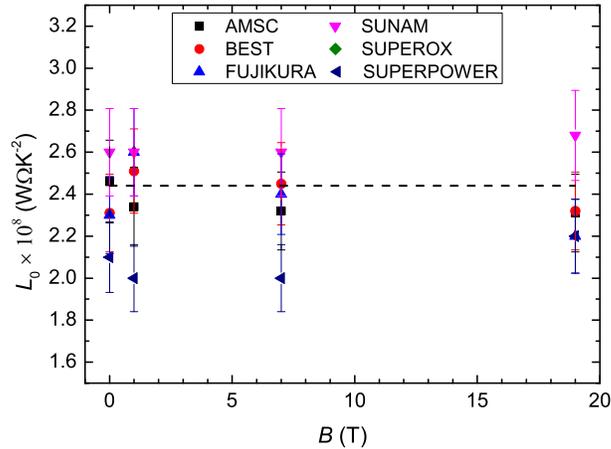} \caption{\label{Lorenz}
Low-temperature value of the transverse Lorenz number of the Cu stabilizer present in the investigated coated conductors. Error bars indicate absolute errors.}
\end{figure}

The present study on thermal properties of REBCO CCs gives the opportunity to evaluate the limits of the procedures commonly adopted for calculating in-field thermal conductivity. In Figure~\ref{Lorenz} we show the field variation of the low-$T$ value of the Lorenz number, $L_0$, in the transverse configuration for all the investigated samples. Since at low temperature $\kappa=(L_0/\rho_{res})T$, we calculated $L_0$ as $\partial\kappa/\partial T \cdot \rho_{res}$, evaluating $\partial\kappa/\partial T$ for $T<10$ K. This procedure allows reducing the error with respect to a direct calculation of the Lorenz number at each temperature.
Contrary to what observed by Arenz \textit{et al.} \cite{Arenz}, our results do not suggest a dependence of the Lorenz number of Cu on the magnetic field in the perpendicular orientation. This confirms the validity of the approximation in Equation~\ref{MR}.

\begin{figure}[t,b]
\centering \includegraphics[width=8 cm]{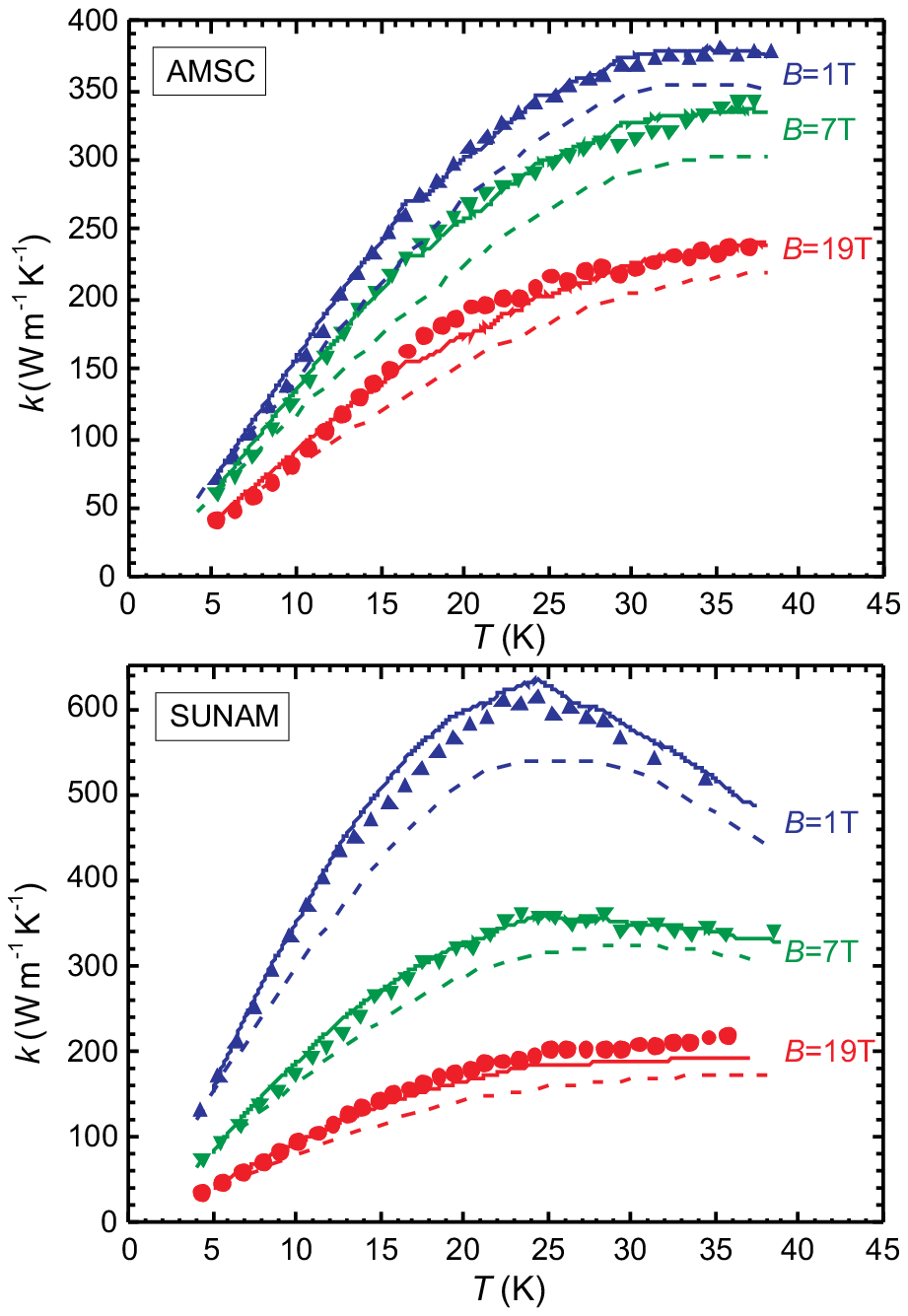} \caption{\label{All_k_3}
Temperature dependence of the thermal conductivity for different values of the magnetic field for the CCs from AMSC and SuNAM. The field has been applied perpendicularly to the thermal-current direction. Points indicate the experimental data. The continuous and dashed lines are calculated values obtained as described in the text.}
\end{figure}

Figure~\ref{All_k_3} presents the experimental in-field $\kappa(T,B)$ data at $B=1$~T, $B=7$~T and $B=19$~T for the CCs produced by AMSC (laminated Cu, low $RRR$) and SuNAM (electroplated Cu, high $RRR$). In the same graph we also report as continuous lines the curves calculated by the formula $\kappa(T,B)=[\rho(T,0)/\rho(T,B)] \kappa(T,0)$, using the measured values for the electrical resistivity and $\kappa(T,0)$. The remarkable agreement between calculated and experimental data further confirms the correctness of the approximation $L(T,B)= L(T,0)$ for magnetic fields up to 19~T. The dotted lines in Figure~\ref{All_k_3} indicate the curves calculated using the $\kappa(T,0)$ values obtained from Equation~\ref{k_Cu} and the $\rho(T,B)$ data as determined from the Kohler formula as described in Reference~\cite{Manfreda}. The largest discrepancy between measured and estimated in-field $\kappa$ values has been observed for the tape from SuNAM and is about $20\%$. However, we would like to remind that the Kohler formula has been deduced starting from measurements performed with the magnetic field perpendicular to the electric current and thus the use of the above-described procedure restricts only to this configuration.

\subsection{Thermal Stability}\label{Discussion-Stability}
Thermal conduction through the superconducting tape is an essential parameter for the thermal stability of a magnet. Due to a local mechanical or thermal disturbance, the current flowing in the tape may exceed the critical current $I_{C}(T^{*})$, $T^{*}$ being the temperature in the region where the perturbation occurs. This implies that only a fraction of the current can flow lossless in the SC whilst the remaining part circulates through the stabilizing metal. In the case of LTS-HTS hybrid magnets, a thermal runaway of the HTS insert can also occur as a consequence of a quench of the LTS winding, due to the inductive coupling of the coils \cite{Maeda,Maeda2}.

Two parameters are very important in the description of quench phenomena in superconductors, namely the minimum quench energy ($MQE$) and the normal zone propagation velocity ($NZPV$) \cite{Wilson}. The first one represents the minimum amount of energy necessary to trigger a quench in the case of a local perturbation. The $NZPV$ describes how fast the overheated zone propagates during a quench.

Starting from the heat diffusion equation, it can be shown that $MQE\propto (\kappa_{Cu}/\rho_{Cu} J_{Cu}^2)^{1/2}$ and $NZPV\propto (\kappa_{Cu} \rho_{Cu} J_{Cu}^2)^{1/2}$, where $J_{Cu}=I_0/S_{Cu}$ is the current density calculated in the approximation that all the magnet current, $I_0$, flows into the Cu as a consequence of the thermal disturbance \cite{Wilson,van Weeren}. The use of pure Cu has the double advantage of increasing $\kappa_{Cu}$ and reducing $\rho_{Cu}$, thus raising the perturbation energy threshold that can be reabsorbed by the system without any quench. By using the Wiedemann-Franz law we deduce: $MQE\propto \kappa_{Cu}/J_{Cu}(LT^{*})^{1/2}=\kappa_{Cu}S_{Cu} /I_0(LT^{*})^{1/2}$. Therefore, at given operating conditions the product $\Pi\equiv \kappa_{Cu}S_{Cu}$ is a relevant parameter for evaluating the thermal stability of the tape. $\Pi$ values calculated at $T\approx 30$~K and $B= 19$~T, with the field applied perpendicularly to the thermal/electrical current, are reported in Table~\ref{TabSamples2}. $\Pi$ varies over more than one order of magnitude, indicating very different thermal stability properties among the examined CCs. In particular, high $\Pi$ values (of the order of 100~$\mu$WK$^{-1}$m) have been found for the tapes produced by AMSC and Fujikura, intermediate values ($\approx 30 \mu$WK$^{-1}$m) for SuNAM and SuperPower, low values ($\approx 10 \mu$WK$^{-1}$m) for BHTS and SuperOx. The highest $\Pi$ value found for the CC from AMSC ($\Pi\approx 110 \mu$WK$^{-1}$m) is mainly a consequence of the large amount of Cu present in this CC ($s_{Cu}\approx0.51$).

From our study we can also deduce that $NZPV$ should not be particularly affected by the magnetic field. In fact, $NZPV$ is proportional to the Lorenz number and we have observed that $L_0$ is field invariant within our experimental accuracy. Low values for $NZPV$ represent a severe problem for HTS-based devices. New layouts for the CC are being investigated in order to improve the quench propagation velocity \cite{Lacroix,Lacroix2}.

For sake of completeness, in Table~\ref{TabSamples2} we have also reported the engineering critical current density, $J_{Eng}\equiv I_C/S_{tot}$, measured at $T=4.2$~K and $B=19$~T with the field applied perpendicularly to the plane of the tape. It is not surprising that $J_{Eng}$ and $\Pi$ are not correlated. However, we would like to point out that thermal properties can be tailored by manufacturers to meet the specific needs of end users. Indeed, stabilization involves processes performed at the last step of the fabrication. Finally, we would like to mention that $I_C$ values of commercial CCs are rapidly evolving as a consequence of the progresses in the fabrication routes. Therefore, $J_{Eng}$ values reported in Table~\ref{TabSamples2} may not represent the best performance achieved by the different manufacturers.

\section{Conclusion}

We have reported on the thermal transport properties of REBCO CCs from six industrial manufacturers. We have shown that zero-field thermal conductivity can be estimated with an accuracy of $\pm15\%$ from the $RRR$ of the Cu stabilizer and the Cu/non-Cu ratio. Thermal conduction has also been investigated in magnetic fields up to 19~T, oriented both parallelly and perpendicularly to the thermal current. Thermal conductivity is reduced on applying the field by a factor that depends on the $RRR$, on the Cu/non-Cu ratio, and on the direction of the field. In particular, larger variations have been observed when $B$ is perpendicular to the thermal-current direction.

Our study has also allowed us to evaluate the consistency of the analytical procedures generally used for estimating in-field thermal conductivity from an electrical characterization of the materials at $B=0$. We have found that the experimental $\kappa$($T,B$) curves in the perpendicular configuration can be estimated with an accuracy of $\pm20\%$ in the framework of the Wiedemann-Franz law. Considerations about the thermal stability of the CCs have been also presented.

\section*{Acknowledgment}
The authors acknowledge the financial support from the Swiss National Science Foundation (Grant N. PP00P2-14467 and 51NF40-144613). Research also supported by FP7 EuCARD-2 http://eucard2.web.cern.ch. EuCARD-2 is co-funded by the partners and the European Commission under Capacities 7th Framework
Programme, Grant Agreement 312453. The authors would like to thank Giorgio Mondonico for the $I_C$ values and more generally for his interest in this work, Lidia Rossi for her support during the in-field $\kappa(T)$ measurements on the Fujikura CC, Christian Barth for the careful reading of the manuscript, Eug\'{e}nie Gallay and Damien Zurmuehle for technical support. The authors also warmly acknowledge Riccardo Tediosi from Bruker Biospin and Giuseppe Celentano from ENEA Frascati for having supplied the samples from Fujikura and SuNAM, and from American Superconductor, respectively.

\end{document}